\documentclass[aps,prl,preprint,nopacs,superscriptaddress]{revtex4}

\usepackage{graphicx}
\usepackage{verbatim}
\usepackage{relsize}
\usepackage{mathrsfs}
\usepackage{color}
\usepackage{colortbl}
\usepackage{ulem}
\usepackage{blindtext}

\usepackage{amsmath,amsfonts,amssymb}
\usepackage{graphicx}

\renewcommand\({\ensuremath \left(}
\renewcommand\){\ensuremath \right)}

\definecolor{mygray}{gray}{0.6}

\def \beq {\begin{equation}}
\def \eeq {\end{equation}}
\begin{document}

\title{Observation of the nonlinear Hall effect under time reversal symmetric conditions\\}

\author{Qiong Ma\footnote{These authors contributed equally to this work.}}\affiliation{Department of Physics, Massachusetts Institute of Technology, Cambridge, Massachusetts 02139, USA}
\author{Su-Yang Xu$^*$}\affiliation {Department of Physics, Massachusetts Institute of Technology, Cambridge, Massachusetts 02139, USA}

\author{Huitao Shen$^*$}\affiliation {Department of Physics, Massachusetts Institute of Technology, Cambridge, Massachusetts 02139, USA}
\author{David Macneill}\affiliation{Department of Physics, Massachusetts Institute of Technology, Cambridge, Massachusetts 02139, USA}
\author{Valla Fatemi}\affiliation{Department of Physics, Massachusetts Institute of Technology, Cambridge, Massachusetts 02139, USA}
\author{Tay-Rong Chang}
\affiliation{Department of Physics, National Cheng Kung University, Tainan 701, Taiwan}

\author{Andr\'es M. Mier Valdivia}\affiliation{Department of Physics, Massachusetts Institute of Technology, Cambridge, Massachusetts 02139, USA}
\author{Sanfeng Wu}\affiliation{Department of Physics, Massachusetts Institute of Technology, Cambridge, Massachusetts 02139, USA}

\author{Zongzheng Du}
\affiliation{Shenzhen Institute for Quantum Science and Engineering and Department of Physics, Southern University of Science and Technology, Shenzhen 518055, China}

\affiliation{Shenzhen Key Laboratory of Quantum Science and Engineering, Shenzhen 518055, China}
\affiliation{School of Physics, Southeast University, Nanjing 211189, China}
\author{Chuang-Han Hsu}
\affiliation{Department of Physics, National University of Singapore, Singapore 117542}
\affiliation{Centre for Advanced 2D Materials and Graphene Research Centre, National University of Singapore, Singapore 117546}

\author{Shiang Fang}
\affiliation{Department of Physics, Harvard University, Cambridge, Massachusetts 02138, USA}
\author{Quinn D. Gibson}
\affiliation{Department of Chemistry, Princeton University, Princeton, New Jersey 08544, USA}
\author{Kenji Watanabe}

\affiliation{National Institute for Materials Science, Namiki 1 -1, Tsukuba, Ibaraki 305 -0044, Japan}
\author{Takashi Taniguchi}

\affiliation{National Institute for Materials Science, Namiki 1 -1, Tsukuba, Ibaraki 305 -0044, Japan}

\author{Robert J. Cava}
\affiliation{Department of Chemistry, Princeton University, Princeton, New Jersey 08544, USA}
\author{Efthimios Kaxiras}
\affiliation{Department of Physics, Harvard University, Cambridge, Massachusetts 02138, USA}\affiliation{John A. Paulson School of Engineering and Applied Sciences, Harvard University, Cambridge, Massachusetts 02138, USA}

\author{Hai-Zhou Lu}
\affiliation{Shenzhen Institute for Quantum Science and Engineering and Department of Physics, Southern University of Science and Technology, Shenzhen 518055, China}
\affiliation{Shenzhen Key Laboratory of Quantum Science and Engineering, Shenzhen 518055, China}

\author{Hsin Lin}
\affiliation{Institute of Physics, Academia Sinica, Taipei 11529, Taiwan}

\author{Liang Fu}\affiliation {Department of Physics, Massachusetts Institute of Technology, Cambridge, Massachusetts 02139, USA}
\author{Nuh Gedik$^{\dag}$}\affiliation {Department of Physics, Massachusetts Institute of Technology, Cambridge, Massachusetts 02139, USA}

\author{Pablo Jarillo-Herrero\footnote{Corresponding authors (emails): gedik@mit.edu and pjarillo@mit.edu }}\affiliation {Department of Physics, Massachusetts Institute of Technology, Cambridge, Massachusetts 02139, USA}

\date{\today}

\maketitle

\textbf{The electrical Hall effect is the production of a transverse voltage under an out-of-plane magnetic field \cite{hall1879new}. Historically, studies of the Hall effect have led to major breakthroughs including the discoveries of Berry curvature and the topological Chern invariants \cite{xiao2010berry, nagaosa2010anomalous}. In magnets, the internal magnetization allows Hall conductivity in the absence of external magnetic field \cite{nagaosa2010anomalous}. This anomalous Hall effect (AHE) has become an important tool to study quantum magnets \cite{nagaosa2010anomalous, chang2013experimental, binz2008chirality,  nakatsuji2015large,yasuda2016geometric, liu2017giant}. In nonmagnetic materials without external magnetic fields, the electrical Hall effect is rarely explored because of the constraint by time-reversal symmetry. However, strictly speaking, only the Hall effect in the linear response regime, i.e., the Hall voltage linearly proportional to the external electric field, identically vanishes due to time-reversal symmetry \cite{ashcroft2005solid}. The Hall effect in the nonlinear response regime, on the other hand, may not be subject to such symmetry constraints \cite{deyo2009semiclassical, moore2010confinement, sodemann2015quantum}. Here, we report the observation of the nonlinear Hall effect (NLHE) \cite{sodemann2015quantum} in the electrical transport of the nonmagnetic 2D quantum material, bilayer WTe$_2$. Specifically, flowing an electrical current in bilayer WTe$_2$ leads to a nonlinear Hall voltage in the absence of magnetic field. The NLHE exhibits unusual properties sharply distinct from the AHE in metals: The NLHE shows a quadratic \textit{I}-\textit{V} characteristic; It strongly dominates the nonlinear longitudinal response, leading to a Hall angle of $\mathbf{\sim90^{\circ}}$. We further show that the NLHE directly measures the ``dipole moment'' \cite{sodemann2015quantum} of the Berry curvature, which arises from layer-polarized Dirac fermions in bilayer WTe$_2$. Our results demonstrate a new Hall effect and provide a powerful methodology to detect Berry curvature in a wide range of nonmagnetic quantum materials in an energy-resolved way. }

\begin{figure*}[t]
\includegraphics[width=16cm]{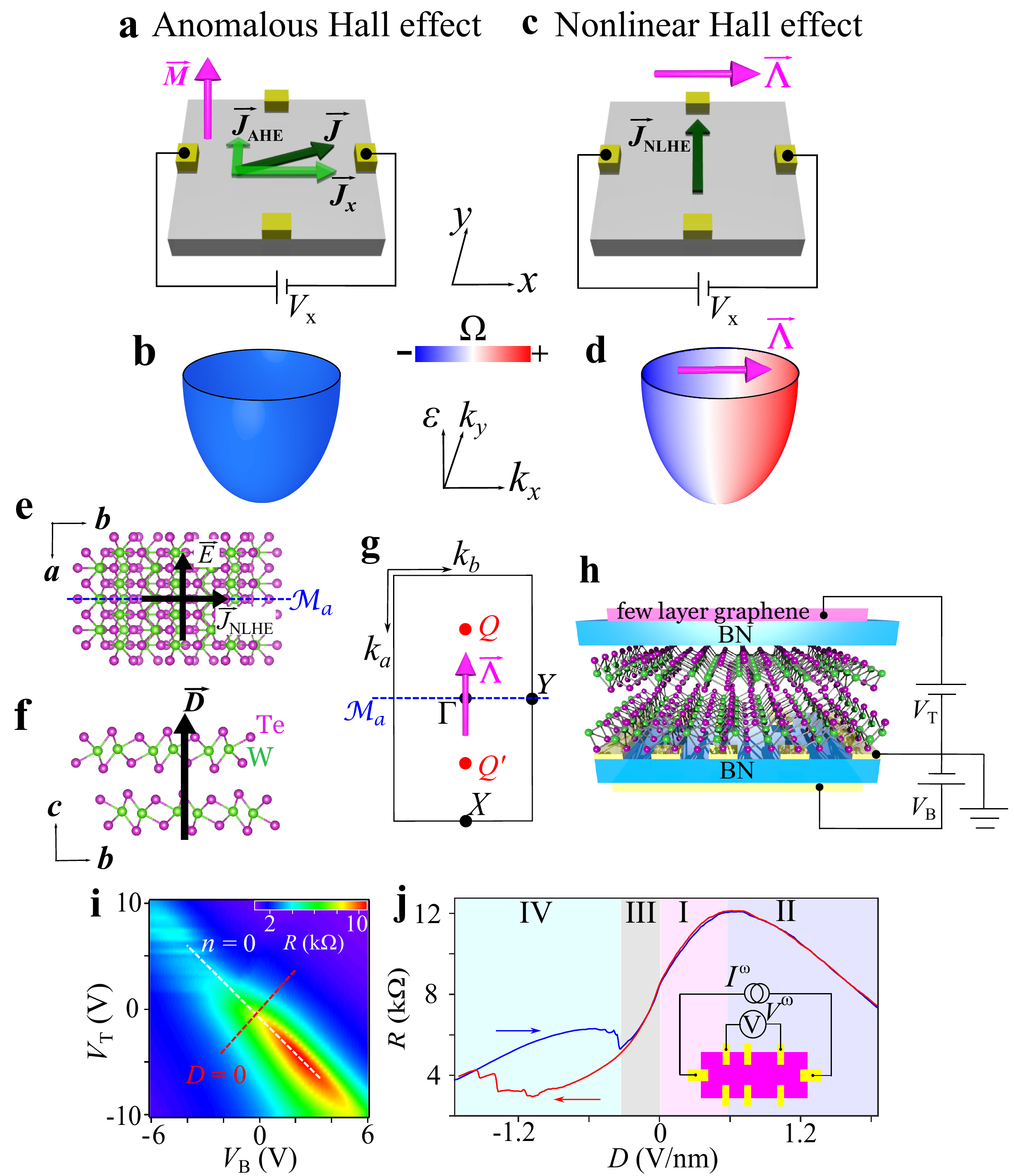}
\caption{{\bf Crystal structure and basic characterization of bilayer WTe$_2$.} \textbf{a,} Illustration of the anomalous Hall effect in a magnetic metal. \textbf{b,} Schematic band structure and Berry curvature}
\label{Fig1}
\end{figure*}

In 1879 Edwin H. Hall observed that, when an electrical current passes through a gold film under a magnetic field, a transverse voltage develops \cite{hall1879new}. This effect, known as the Hall effect, forms the basis of both fundamental research and practical applications such as magnetic field measurements and motion detectors. In contrast to the classical Hall effect where the Lorentz force bends the trajectory of the charge carriers, quantum mechanics describes the ``bending'' by the intrinsic geometry of the quantum electron wavefunctions under time-reversal symmetry breaking. This crucial theoretical understanding eventually led to the seminal discoveries of the Berry curvature and the topological Chern number, which have become pillars of modern condensed matter physics \cite{xiao2010berry, nagaosa2010anomalous}. One important current frontier is to identify AHE with quantized or topological character in unconventional magnetic quantum materials, where spin-orbit coupling (SOC), geometrical frustration and electronic correlations coexist \cite{nagaosa2010anomalous, chang2013experimental, binz2008chirality, nakatsuji2015large, yasuda2016geometric, liu2017giant}. These extensive studies \cite{hall1879new, nagaosa2010anomalous, chang2013experimental,  binz2008chirality, nakatsuji2015large, yasuda2016geometric, liu2017giant} have established a paradigm for the electrical Hall effect: (1) A non-vanishing Hall conductivity arises from the momentum-integrated Berry curvature. This requires the breaking of time-reversal symmetry and therefore is realized in magnets or by applying magnetic fields (Figs.~\ref{Fig1}\textbf{a,b}); (2) The Hall voltage is linearly proportional to the external electric field; (3) The Hall conductivity is usually a fraction of the longitudinal conductivity (except in the quantum Hall regime), as measured by the Hall angle. 

However, electrical Hall effects beyond this paradigm are in fact theoretically possible. While Hall effects in the linear response regime (abbreviated as LHE) must exhibit the properties described above, those in the nonlinear response regime (abbreviated as NLHE) are not subject to the same constraints and therefore may possess completely distinct characteristics. Building upon previous theoretical works \cite{deyo2009semiclassical, moore2010confinement}, the second-order NLHE was recently proposed in Ref. \cite{sodemann2015quantum}. It was pointed out that \cite{sodemann2015quantum}, apart from the momentum-integrated Berry curvature that is responsible for the LHE, there can be other important properties concerning the Berry curvature. Particularly, even in a nonmagnetic material, inversion symmetry breaking may segregate the positive and negative Berry curvatures at different $k$ regions (Fig.~\ref{Fig1}\textbf{d}), leading to a dipole moment. Such a ``Berry curvature dipole'' was proposed only very recently \cite{sodemann2015quantum} and its various experimental consequences are still being explored \cite{sodemann2015quantum, lee2017valley, xu2018electrically}. Interestingly, the Berry curvature dipole can also give rise to an electrical Hall effect \cite{sodemann2015quantum}, albeit in the second-order response. As depicted in Fig.~\ref{Fig1}\textbf{c}, a nonlinear Hall effect is induced by an electric field parallel to the Berry curvature dipole $\vec{\Lambda}$. The exciting possibility of realizing electrical Hall effects in a wide class of nonmagnetic materials immediately attracted great interest \cite{low2015topological, morimoto2016chiral, zhang2017berry, tsirkin2018gyrotropic, zhang2018electrically, you2018berry, shi2018berry, Facio2018}. We have carefully considered possible material platforms according to the following three criterions: the existence of Berry curvature hotspots, the desired crystalline symmetry properties and the presence of additional experimental tuning parameters. We have identified the 2D quantum material, bilayer WTe$_2$, as an ideal material platform (see supplementary information SI.I).

Recently, WTe$_2$ has attracted significant interest because of its remarkable properties both in bulk and in monolayer \cite{ali2014large, soluyanov2015type, Qian2014, macneill2016, Zheng2016, Fei2017, Tang2017, wu2018observation, fatemi2018gate, Cobdentalk}. In particular, monolayer WTe$_2$ was found to show a high-temperature quantum spin Hall state \cite{Qian2014, Zheng2016, Fei2017, Tang2017, wu2018observation} and gate-tunable superconductivity \cite{fatemi2018gate, Cobdentalk}. On the other hand, the electronic properties of bilayer WTe$_2$ remain relatively unexplored, apart from recent transport measurements \cite{Fei2017} and first-principles calculations \cite{Zheng2016} that showed a semiconductor/semimetal state with a tiny band gap of a few meV. Since monolayer WTe$_2$ features a Dirac fermion at each $Q$($Q'$) point \cite{Qian2014, Fei2017} (Extended Data Fig. 1), bilayer WTe$_2$ can be understood as a pair of coupled Dirac fermions (one from each layer). The presence of Dirac fermions suggests large Berry curvatures if relevant crystalline symmetries are broken \cite{yao2008valley}. Indeed, what makes the bilayer's electronic structure more unusual is its very low crystalline symmetry (Figs.~\ref{Fig1}\textbf{e,f}) \cite{mar1992metal, beams2016characterization, macneill2016, macneill2017thickness}. Specifically, while monolayer WTe$_2$ is (approximately) inversion symmetric \cite{Qian2014,  Zheng2016, xu2018electrically}, bilayer WTe$_2$ is found to be strongly inversion symmetry breaking (Figs.~\ref{Fig1}\textbf{e,f}) \cite{beams2016characterization, macneill2016, macneill2017thickness} because of the stacking arrangement \cite{mar1992metal}, a behavior opposite to the hexagonal transition-metal dichalcogenides (TMD). In fact, the only crystalline symmetry for bilayer WTe$_2$ is the mirror plane $\mathcal{M}_a$ (Fig.~\ref{Fig1}\textbf{e}). Because of $\mathcal{M}_a$, the Berry curvature dipole (a pseudovector) is required to be parallel to the $\hat{a}$ axis (Fig.~\ref{Fig1}\textbf{g}). Therefore, according to the schematic illustration in Fig.~\ref{Fig1}\textbf{c}, a nonlinear Hall voltage is expected to appear along the $\hat{b}$ axis in response to an external electric field along $\hat{a}$ (Fig.~\ref{Fig1}\textbf{e}).

We have fabricated high-quality, encapsulated, dual-gated bilayer WTe$_2$ devices (Fig.~\ref{Fig1}\textbf{h}), which allow us to independently control the charge density $n$ and the out-of-plane electrical displacement field $\mathbf{D}$. To optimize the devices for testing the NLHE, flakes with long, straight edges (an indication of the crystalline axis \cite{beams2016characterization}) were selected and contacts were aligned along the straight edges. We start by studying the four-probe resistance of bilayer WTe$_2$. Remarkably, the resistance is strongly asymmetric about $D=0$ (Fig.~\ref{Fig1}\textbf{i}). To clearly illustrate this property, we plot the resistance at charge neutrality as a function of $D$ (Fig.~\ref{Fig1}\textbf{j}): From $D=0$ to $D>0$, the resistance increases in regime I and then decreases in regime II. By contrast, from $D=0$ to $D<0$, the resistance decreases in regime III and shows a large hysteresis in regime IV. The observed strong asymmetry in the resistance data in bilayer WTe$_2$ directly arises from the lack of any symmetry that relates $\hat{c}$ to $-\hat{c}$ (see Extended Data Fig. 1 and SI.II.3). Therefore, this is a unique property of bilayer WTe$_2$, which is not present in other known 2D materials such as bilayer graphene, bilayer hexagonal TMDs and monolayer WTe$_2$. The large hysteresis in regime IV suggests a $D$-field driven ferroelectric switching \cite{Cobdentalk}. In SI.II.3, we show that both the asymmetric resistance profile and the ferroelectric switching are evidence for bilayer WTe$_2$'s low symmetry nature, in agreement with previous studies \cite{beams2016characterization, macneill2016, macneill2017thickness}. 

To further confirm the bilayer's inversion symmetry breaking nature and to determine the crystalline axes, we study the circular photogalvanic effect (CPGE) in the mid-infrared regime ($\hbar\omega\simeq120$ meV). As shown in Fig.~\ref{Fig2}\textbf{c}, the CPGE in bilayer WTe$_2$ is about an order of magnitude stronger than that of monolayer WTe$_2$, which clearly demonstrates the bilayer's strong inversion symmetry breaking. Moreover, the $D$-field dependence of the CPGE is asymmetric about $D=0$ in bilayer but roughly symmetric in monolayer (Fig.~\ref{Fig2}\textbf{c}), consistent with the resistance behavior (Figs.~\ref{Fig1}\textbf{i,j}). Furthermore, the CPGE's clear directional dependence allows us to determine the crystalline axes (SI.IV): As shown in Fig.~\ref{Fig2}\textbf{b}, the direction connecting electrodes $7-8$ roughly corresponds to the crystalline $\hat{a}$ axis, which is also consistent with the shape of the bilayer flake (Fig.~\ref{Fig2}\textbf{a}).

With the inversion symmetry breaking nature confirmed and the crystalline axes determined, we now explore the NLHE in bilayer WTe$_2$. Phenomenologically, the NLHE can be described as $V^{\textrm{NLHE}}_{\alpha\beta\beta}\propto E_\beta E_\beta$, where $E$ is the external electric field, $\alpha$ and $\beta$ are the in-plane spatial directions ($\alpha \perp \beta$ for the Hall effect), and $V^{\textrm{NLHE}}_{\alpha\beta\beta}$ is the nonlinear Hall voltage along $\alpha$ in response to the external electric field along $\beta$. Hence, an AC electric field with frequency $\omega$ should generate a voltage with double frequency $V^{2\omega}$ \cite{sodemann2015quantum}, which can be directly measured by a lock-in amplifier in a phase-sensitive way. We study the essential, qualitative behavior of $V^{2\omega}$ at $T=50$ K. By keeping the AC excitation current $I^{\omega}_{\textrm{exc}}$ parallel to $\hat{a}$, we test both the longitudinal and transverse $V^{2\omega}$ (Fig.~\ref{Fig2}\textbf{d}): Indeed, we observe a clear transverse $V^{2\omega}$ response between electrodes $3-4$ ($V^{2\omega}_{baa}$), which strongly dominates over the longitudinal response between electrodes $1-3$ and $1-5$. Moreover, the transverse $V^{2\omega}$ shows a clear quadratic $I-V$ characteristic, which, combined with the $2\omega$ frequency, establishes its second-order nature. In addition, the transverse $V^{2\omega}_{baa}$ signal shows a strong gate voltage dependence (Fig.~\ref{Fig2}\textbf{e}), while the quadratic $I-V$ characteristic remains robust at all gate voltages.

We then perform systematic measurements at $T=10$ K (Fig. 3). The gate map of $V^{2\omega}_{baa}$ (Fig.~\ref{Fig3}\textbf{a}) exhibits a strong dependence on both the charge density and displacement field. The black dashed line in Fig.~\ref{Fig3}\textbf{a} tracks the ($V_{\textrm{T}}, V_{\textrm{B}}$) locations, which correspond to the maximum of the resistance (Fig.~\ref{Fig1}\textbf{i}) at every $D$-field value (SI.II.1). We observe the following features in Fig.~\ref{Fig3}\textbf{a}: (1) At $D=0$, $V^{2\omega}_{baa}$ shows the same sign on the opposite sides of the resistance peak (along the charge density axis). Upon further hole doping, the sign is reversed. (2) By applying a sufficiently large positive displacement field ($D>0$), an additional sharp blue feature emerges very close to the charge neutrality in between the two red features. In Fig.~\ref{Fig3}\textbf{b}, we compare the charge density dependence of $V^{2\omega}_{baa}$ and resistance along the green dashed line in Fig.~\ref{Fig3}\textbf{a}. The resistance is tightly related to properties derived from the $\varepsilon-k$ dispersion (e.g. the effective mass, the Fermi velocity, etc) and the defect scattering time $\tau$. Therefore, the observed typical ambipolar behavior shows that the above properties evolve smoothly as a function of charge density. On the other hand, $V^{2\omega}_{baa}$ exhibits multiple sign reversals along the same trajectory. The contrasting behavior reveals a distinct physical origin for $V^{2\omega}_{baa}$ beyond the basic $\varepsilon-k$ dispersion properties and the defect scattering time $\tau$: $V^{2\omega}_{baa}$ arises from a property whose sign depends sensitively on the chemical potential $\varepsilon_{\textrm{F}}$. In Extended Data Fig. 3, we further show the temperature dependence of $V^{2\omega}_{baa}$ and resistance. In Fig.~\ref{Fig3}\textbf{c}, we show the measured $V^{2\omega}$ signals between many combinations of electrodes. Independent of the choice of electrodes, the transverse response is strongly dominant over the longitudinal response. Equally importantly, all transverse $V^{2\omega}_{baa}$ data curves measured between the many combinations of electrodes (Fig.~\ref{Fig3}\textbf{c}) collapse onto each other. These measurements clearly demonstrate the Hall nature of the $V^{2\omega}$ signals. Our data in Fig.~\ref{Fig3}\textbf{c} achieves a remarkable Hall angle of $\sim90^{\circ}$ for the nonlinear response (Fig.~\ref{Fig3}\textbf{d}). 

We enumerate here the key essential data, including the clear observation of the nonlinear Hall voltage $V^{2\omega}_{baa}$, its dominance over the longitudinal response, its unique gate dependence, and its consistency between many combinations of electrodes. These data are crucial for excluding various extrinsic effects.  For instance, an accidental diode at the contact/sample interface can lead to a rectification effect. However, that extrinsic signal should not show ``Hall dominance'', and can also be ruled out by the consistent data between many combinations of electrodes. In SI.VIII, we present a systematic discussion on how extrinsic effects are carefully considered and excluded. We now consider the Berry curvature dipole induced NLHE as the possible origin for our data. Indeed, our observation of the nonlinear Hall voltage $V^{2\omega}_{baa}$ is consistent with the general symmetry expectation for the Berry curvature dipole effect in bilayer WTe$_2$ (Fig.~\ref{Fig1}\textbf{e}). To more directly confirm this origin, we take advantage of the additional tuning parameters enabled by the dual gating. In particular, our $V^{2\omega}_{baa}$ data exhibits a unique gate dependence, consisting of a strong dependence on the displacement $D$-field and multiple sign-reversals as a function of the chemical potential $\varepsilon_\textrm{F}$. These distinct experimental features over a large parameter space provide rich information about the physical origin, which can be directly compared with the Berry curvature behavior in bilayer WTe$_2$. 

We provide a physical picture to elucidate the highly unusual Berry curvature dipole in bilayer WTe$_2$. This physical picture allows us to understand why a large Berry curvature dipole can be formed by asymmetrically coupled Dirac fermions and some of its key features at different energies. \color{black} As a starting point, a massless Dirac fermion (as in graphene) has no Berry curvature (Fig.~\ref{Fig4}\textbf{a}). Gapping it out by breaking inversion symmetry (as in gapped graphene or hexagonal TMD monolayers) leads to large Berry curvatures $\Omega$ near the gap edges \cite{yao2008valley} (Fig.~\ref{Fig4}\textbf{b}). The Berry curvature dipole $\vec{\Lambda}$, however, is still zero, because $\Omega$ is uniform around the Fermi surface (Fig.~\ref{Fig4}\textbf{b}). To describe the same fact mathematically, we write down the expression for $\vec{\Lambda}$ \cite{sodemann2015quantum}:

\begin{equation}
\Lambda_\alpha=\frac{1}{\hbar}\int d^2\mathbf{k} \delta(\varepsilon-\varepsilon_{\textrm{F}}) \frac{\partial\varepsilon}{\partial k_\alpha} \Omega(\mathbf{k}),
\label{BCD}
\end{equation}

where $\varepsilon$ and $\mathbf{k}$ are the energy and wave-vector, $\varepsilon_{\textrm{F}}$ is the Fermi energy, and $\frac{\partial\varepsilon}{\partial k_\alpha}$ is the band slope along $\alpha$. This integral vanishes because $\Omega$ is a constant whereas $\frac{\partial\varepsilon}{\partial k_\alpha}$ is equal but opposite on the opposite sides of the Fermi surface. Tilting the Dirac cone allows the left- and right-movers to have different $\Omega$ and different $\frac{\partial\varepsilon}{\partial k_\alpha}$ (Fig.~\ref{Fig4}\textbf{c}), leading to a nonzero $\vec{\Lambda}$. Importantly, for a single Dirac fermion, $\vec{\Lambda}$ is opposite for the conduction and valence bands (see Fig.~\ref{Fig4}\textbf{c} and Eq.~\ref{BCD}). We now consider bilayer WTe$_2$. In the absence of inter-layer coupling and SOC, the $Q$ point features a pair of decoupled, massless Dirac fermions of opposite chirality (one from each layer, Fig.~\ref{Fig4}\textbf{d}). We now turn on inter-layer coupling: (1) The two Dirac fermions are both gapped because of the explicit breaking of inversion symmetry due to inter-layer coupling (Fig.~\ref{Fig4}\textbf{e}). (2) Also because of the coupling, the two Dirac fermions are shifted with respect to each other and open up gaps wherever they cross (Fig.~\ref{Fig4}\textbf{f}). To capture the essential physics, we only considered the shift along momentum but ignored the shift along energy. As a result, we arrive at a quite dramatic scenario where the left- and right-movers have the opposite Berry curvatures (Fig.~\ref{Fig4}\textbf{f}). This relatively simple physical picture already captures some of the key features in our data: (1) The asymmetrically coupled Dirac fermions lead to a large Berry curvature dipole along $\hat{a}$, consistent with the observed nonlinear Hall voltage $V_{baa}^{2\omega}$; (2) The Berry curvature dipole has the same sign for the conduction and valence bands immediately next to charge neutrality, consistent with the $V_{baa}^{2\omega}$ data at $D=0$. In SI.V and VI, we further consider the shift along energy, the inclusion of SOC and the effect of the $D$-field, which allow us to capture and understand other main features of the data including the sign reversal at large hole dopings and the appearance of the sharp blue feature. In particular, the sharp blue feature at $D>0$ (Fig.~\ref{Fig3}\textbf{a}) arises from an anti-crossing (a massive Dirac fermion) formed between the lowest two conduction bands (the green circle in Extended Data Fig. 2\textbf{d}, see SI.V).

\color{black} Finally, we determine the Berry curvature dipole $\Lambda_a$ (Figs.~\ref{Fig4}\textbf{g-i}) based on the $V^{2\omega}_{baa}$ data (SI. VII) and directly compare with the first-principles calculated results. At $D=0$ and $D=0.6$ V/nm (Figs.~\ref{Fig4}\textbf{g,h}), $\Lambda_a$ shows a positive peak on either side of $\varepsilon_\textrm{F}=0$; Moving further away from $\varepsilon_\textrm{F}=0$, $\Lambda_a$ turns negative. These behaviors are consistent with the calculations in Figs.~\ref{Fig4}\textbf{j,k}. At $D=1.6$ V$/$nm, a sharp, negative feature appears in the data near $\varepsilon_\textrm{F}=0$ (Fig.~\ref{Fig4}\textbf{i}), which is also captured in theory (Fig.~\ref{Fig4}\textbf{l}); In fact, calculations show two sharp peaks immediately next to each other. The positive one merges with a broader feature of the same sign. The agreement between data and first-principle calculations provide compelling evidence for our experimental observation of the Berry curvature dipole induced NLHE.

%\color{red}
Our results demonstrate a new type of Hall effect. Remarkably, this is an electrical Hall effect in a nonmagnetic material and in the absence of magnetic field. We also highlight its importance as a Berry curvature probe. While the AHE has been widely used to measure the Berry curvature of magnetic metals/semimetals, a similar probe for nonmagnetic counterparts has remained elusive. Importantly, the NLHE signal always corresponds to the Berry curvature dipole at the Fermi level, independent of band structure details (e.g. the absence/presence of SOC and the number of Fermi surfaces); To get the Berry curvature dipole from the NLHE signal, only the Drude relaxation time $\tau$ is further needed, which can be obtained from standard electrical transport. These characteristics highlight the NLHE as a powerful and universal probe of the Berry curvatures in nonmagnetic quantum metals and semimetals. This is complementary to previous optoelectronic approaches in wide-gap semiconductors with valley selection rules \cite{yao2008valley, mak2014valley}. Moreover, the NLHE provides an important inspiration: How Berry curvature distribute in $k$ space (e.g. the Berry curvature dipole and Berry curvature quadruple) can modify electrical, thermoelectric, optical, plasmonic and other key properties of quantum materials, giving rise to novel effects beyond existing paradigm. More broadly, nonlinear electrical transport from intrinsic quantum or topological properties is an unexplored territory with ample exciting possibilities. For instance, many ordered states (charge-density waves, superconductivity, etc.) manifest as the breaking of a particular symmetry for the electrons near the Fermi level. Because the nonlinear transport here combines sensitivity to symmetry as in nonlinear optics and sensitivity to Fermi energy physics as in regular transport, it may be used to probe order parameters of novel broken symmetry states \cite{qin2017superconductivity, harter2017parity}. Finally, the responsivity of our WTe$_2$ device, defined as the ratio between the output nonlinear voltage and the input power ($\frac{V^{\textrm{NLHE}}}{I^2R}$) reaches a large value of $10^4$ V/W (see SI. IX for discussions). The large responsivity highlights the potential for using intrinsic quantum properties of homogenous quantum materials for nonlinear applications including frequency-doubling and rectification. The generalization to gigahertz or terahertz frequencies might be useful for next generation wireless technologies.

\color{black}

\bibliographystyle{naturemag}
\bibliography{Topological_and_2D_12062017}

\vspace{0.5cm}
\textbf{Methods}

\textbf{Sample fabrication:} Our fabrication of the dual-gated bilayer WTe$_2$ devices consists of two phases. Phase I was done under ambient conditions: local bottom PdAu or graphite gates were first defined on standard Si/SiO$_2$ substrates. A suitable hexagonal hexagonal BN (hBN) flake ($\sim 10 - 20$ nm thick) was exfoliated onto a separate Si/SiO$_2$ substrate, picked up using a polymer-based dry transfer technique and placed onto the pre-patterned local bottom gate. Electrical contacts (PdAu, $\sim 20$ nm thick) were deposited onto the bottom hBN flake with $e$-beam lithography and metal deposition. Alternatively, patterned graphene fingers were transferred on top of the hBN as the bottom electrodes (for the device in the SI.II.2). Phase II was done fully inside a glovebox with argon environment. Bilayer WTe$_2$ flakes were exfoliated from a bulk crystal onto a Si/SiO$_2$ chip. Few layer graphene, hBN ($\sim 10 - 20$ nm thick) and bilayer WTe$_2$ were sequentially picked up and then transferred onto the local bottom gate/hBN/contact substrate. Extended leads connecting the top gate graphene to wire bonding pads were pre-made together with the metal contacts in Phase I. In such a dual-gated device, the charge density can be obtained by $n=\frac{\epsilon_0\epsilon^{\textrm{hBN}}}{e}(V_{\textrm{T}}/h_{\textrm{T}}+V_{\textrm{B}}/h_{\textrm{B}}$). The displacement field is determined by $D=\epsilon^{\textrm{hBN}}(V_{\textrm{B}}/h_{\textrm{B}}-V_{\textrm{T}}/h_{\textrm{T}})/2$ \cite{taychatanapat2010}. Here $n$ is the charge density, $D$ is the displacement field, $V_{\textrm{T}}$ ($V_{\textrm{B}}$) are the gate voltages, $\epsilon_0=8.85\times10^{-12}$ F/m is the vacuum permittivity and $\epsilon^{\textrm{hBN}} \sim 3$, $h_{\textrm{T}}=16$ nm and $h_{\textrm{B}}=10$ nm are relative dielectric constant, thicknesses of the top ($T$) and bottom ($B$) hBN layers of the presented device, respectively.

\textbf{Mid-infrared scanning photocurrent microscopy:} The laser source is a temperature-stablized CO$_2$ laser ($\lambda = 10.6$ $\mu$m or $\hbar\omega = 120$ meV). A focused beam spot (beam waist $\simeq 25$ $\mu$m) is scanned (using a two axis piezo-controlled scanning mirror) over the entire sample and the current is recorded at the same time to form a two-dimensional map of photocurrent as a function of spatial positions. Reflected light from the sample is collected to form a simultaneous reflection image of the sample. The absolute location of the photo-induced signal is therefore found by comparing the photocurrent map to the reflection image. The light polarization is modulated by a rotatable quarter-wave plate. 

\textbf{Transport measurements:} Electrical transport was measured in a He-$3$ cryostat. The top and bottom gate voltages were applied through two Keithley $2400$ SourceMeters. Both first- and second-harmonic signals were collected by standard Lock-in techniques (Stanford Research Systems Model SR830) with excitation frequencies between $10-200$ Hz. Three frequencies ($17.777$ Hz, $77.77$ Hz, and $177.77$ Hz) were carefully tested, which gave consistent data. The phase of the first- (second-) harmonic signal was approximately $0^{\circ}$ ($\pm 90^{\circ}$), consistent with the expected values for first and second-order responses \cite{qin2017superconductivity, rectification2017, wakatsuki2017}. Specifically, in the presence of an AC excitation current $I_{\textrm{exc}}^\omega=I_0\sin\omega t$, the second-order voltage is expressed as $V^{\textrm{second-order}}\propto (I_0\sin\omega t)^2 = \frac{1}{2} I_0^2 \(1+ \sin(2\omega t-\frac{\pi}{2}) \)$, directly revealing the phase difference between first- and second-harmonic signals. This equation also shows that $V^{\textrm{second-order}}$ always consists of a second-harmonic component and a DC component. The coexistence of second-harmonic and DC components is a generic property for various second-order transport effects including the NLHE \cite{sodemann2015quantum}, the nonreciprocal current with magnetic fields \cite{qin2017superconductivity, rectification2017, wakatsuki2017}, the spin-torque in magnetic materials/hetereostructures \cite{kim2013, garello2013, macneill2016, macneill2017thickness} and for other rectification effects on diodes and other external junctions (see SI. VIII). Similar to measurements on the reciprocal current and the magnetic spin-torque \cite{qin2017superconductivity, rectification2017, wakatsuki2017, kim2013, garello2013, macneill2016, macneill2017thickness}, we focus on the second-harmonic signals because it allows us to utilize the lock-in technique which greatly enhances the measurement sensitivity and precision.

\color{black}
\textbf{First-principles calculations:} First-principles band structure calculations were performed using the projector augmented wave method as implemented in the VASP package \cite{VASP} within the generalized gradient approximation (GGA) schemes. 11$\times$15$\times$1 Monkhorst-Pack $k$-point meshes with an energy cutoff of $400$ eV were used for slab calculations. The vacuum thickness was larger than 20 $\textrm{\AA}$ to ensure the separating of the slabs. In order to correct the energy band gaps, we also performed calculations with the Heyd-Scuseria-Ernzerhof (HSE) hybrid functional \cite{HSE}. We used W $d$ and Te $p$ orbitals to construct Wannier functions \cite{wannier} without performing the procedure for maximizing localization.

\vspace{0.5cm}

\textbf{Data availability:} The data that support the plots within this paper and other findings of this study are available from the corresponding author upon reasonable request.

\textbf{Acknowledgement:} We thank Joe Checkelsky, Feng Qin, Yoshihiro Iwasa, Jhih-Shih You and Inti Sodemann for important discussions. Work in the PJH group was partly supported by the Center for Excitonics, an Energy Frontier Research Center funded by the US Department of Energy (DOE), Office of Science, Office of Basic Energy Sciences (BES) under Award Number DESC0001088 (fabrication and measurement) and partly through AFOSR grant FA9550-16-1-0382 (data analysis), as well as the Gordon and Betty Moore Foundation's EPiQS Initiative through Grant GBMF4541 to PJH. This work made use of the Materials Research Science and Engineering Center Shared Experimental Facilities supported by the National Science Foundation (NSF) (Grant No. DMR-0819762). NG and SYX acknowledge support from DOE, BES DMSE (data taking and analysis), and the Gordon and Betty Moore Foundations EPiQS Initiative through Grant GBMF4540 (manuscript writing). The WTe$_2$ crystal growth performed at Princeton University was supported by an NSF MRSEC grant, DMR-1420541 (QDG and RJC).  ZZD and HZL was supported by Guangdong Innovative and Entrepreneurial Research Team Program (No. 2016ZT06D348), the National Key R \& D Program (No. 2016YFA0301700), the National Natural Science Foundation of China (No. 11574127), and the Science, Technology, and Innovation Commission of Shenzhen Municipality (No. ZDSYS20170303165926217). KW and TT acknowledge support from the Elemental Strategy Initiative conducted by the MEXT, Japan, JSPS KAKENHI Grant Numbers JP18K19136 and the CREST (JPMJCR15F3), JST. HS, LF, and SF acknowledge support from NSF Science and Technology Center for Integrated Quantum Materials grant DMR-1231319 (theory for HS and LF and calculations for SF). TRC was supported by the Ministry of Science and Technology under MOST Young Scholar Fellowship: MOST Grant for the Columbus Program NO. 107-2636-M-006-004-, National Cheng Kung University, Taiwan, and National Center for Theoretical Sciences (NCTS), Taiwan. EK acknowledges support by ARO MURI Award W911NF-14-0247. The computational work at Harvard University was performed on the Odyssey cluster supported by the FAS Division of Science, Research Computing Group.
 
\textbf{Author contributions:} PJH and NG supervised the project. SYX and QM conceived the experiment. QM and SYX performed photocurrent measurements and analysed the data. QM, SYX and DM performed transport measurements with the help from VF. SYX, QM, VF and DM performed data analysis and discussed the results. AMMV, QM, DM, SW and VF fabricated the devices. TRC, CHH and SF calculated the first-principles band structures and the Berry curvature dipole in bilayer WTe$_2$ under the supervision of EK and HL. HS, SYX, ZD performed theoretical modeling and analyses under the supervision of HZL and LF. QDG and RJC grew the bulk WTe$_2$ single crystals. KW and TT grew the bulk hBN single crystals. SYX, QM, NG and PJH wrote the manuscript with input from all authors.

\textbf{Competing financial interests:} The authors declare no competing financial interests.

\clearpage
\addtocounter{figure}{-1}
\begin{figure*}[t!]
\caption{($\Omega$) distribution of a simple magnetic metal. The anomalous Hall conductivity arises from the momentum-integrated Berry curvature, which requires the breaking of time-reversal symmetry. \textbf{c,} Illustration of the nonlinear Hall effect. Because of the in-plane Berry curvature dipole $\vec{\Lambda}$, an electrical bias parallel to the Berry curvature dipole leads to a nonlinear Hall voltage. The linear longitudinal current ($\vec{J}_x$) for conductors is not drawn in this panel. \textbf{d,} Schematic band structure and Berry curvature distribution of a wide class of nonmagnetic, inversion breaking quantum materials with a nonzero Berry curvature dipole. The Berry curvature dipole originates from the segregation of positive and negative Berry curvatures in $k$ space, which does not need to break time-reversal symmetry. \textbf{e,f,} Crystal structure of bilayer WTe$_2$. The only crystalline symmetry is the mirror plane $\mathcal{M}_a$. Because of $\mathcal{M}_a$, the Berry curvature dipole $\vec{\Lambda}$ (a pseudovector) is required to be parallel to $\hat{a}$ in bilayer WTe$_2$. As a result, consulting the schematic illustration in panel (\textbf{c}), in bilayer WTe$_2$, an external electric field along $\hat{a}$ can lead to a nonlinear Hall voltage along $\hat{b}$. \textbf{g,} The first Brillouin zone with important momenta denoted. The pink arrow represents the Berry curvature dipole in bilayer WTe$_2$. \textbf{h,} Schematic illustration of an encapsulated, dual-gated bilayer WTe$_2$ device. \textbf{i,} The four-probe longitudinal resistance $R$ as a function of both top ($V_\mathrm{T}$) and bottom gates ($V_\mathrm{B}$)  at $T=10$ K. The white and red lines correspond to the charge neutrality $n=0$ and the zero displacement field $D=0$, respectively. \textbf{j,} The four-probe resistance at charge neutrality as a function of the displacement field (along the white dashed lines in panel(\textbf{i})). The blue and red curves correspond to the forward and backward displacement field sweeps, respectively. The four regimes (I-IV) of the resistance are denoted. }
\end{figure*}

\clearpage
\begin{figure*}[t]
\includegraphics[width=17cm]{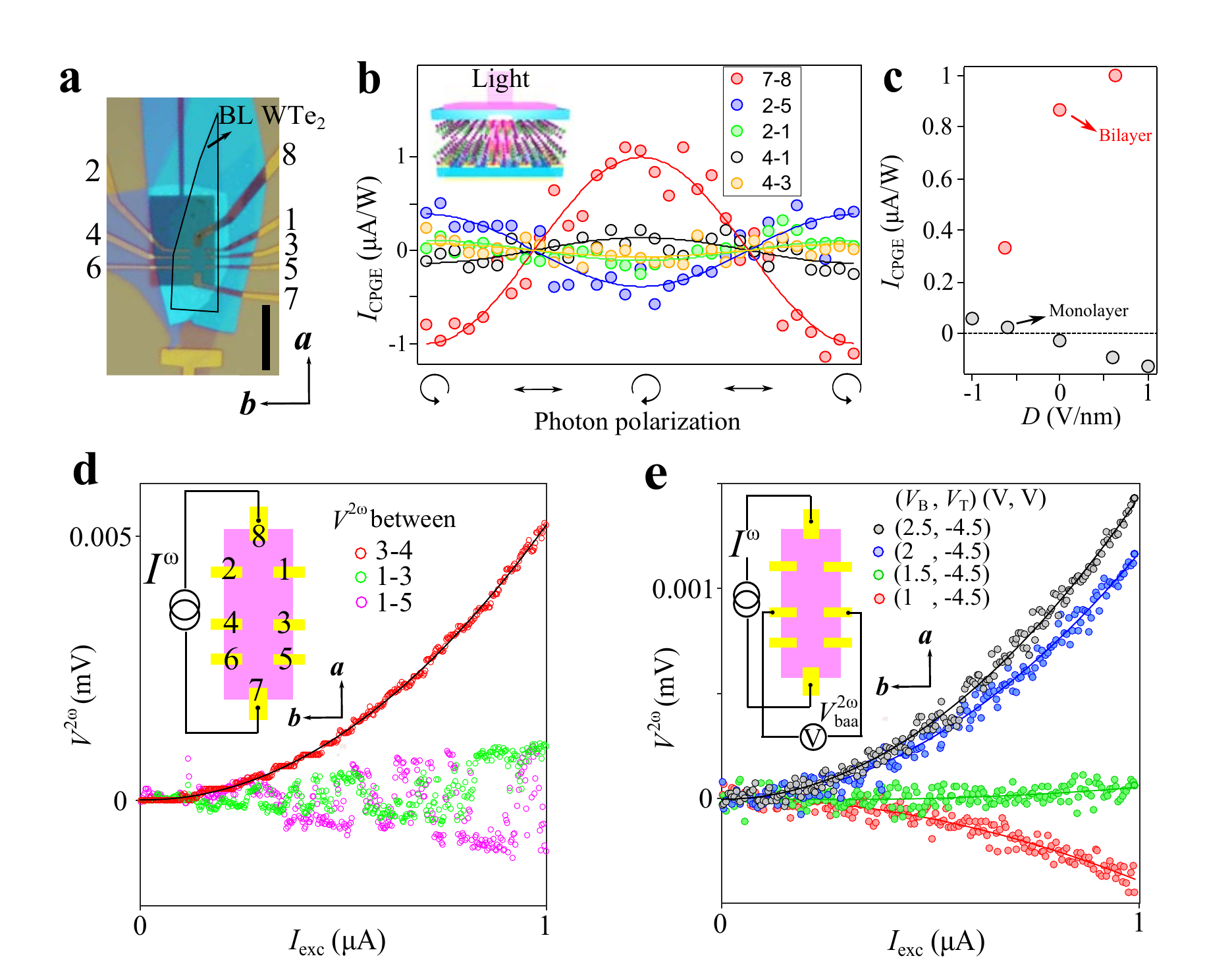}
\caption{{\bf Circular photocurrents and nonlinear transport of bilayer WTe$_2$.} \textbf{a,} Optical image of our WTe$_2$ device with the electrodes numbered. The black lines trace the encapsulated bilayer WTe$_2$ flake. The crystalline directions determined by the circular photogalvanic effect (CPGE) are noted. Scale bar: $15 \ \mu$m. \textbf{b,} Polarization dependent photocurrents between different electrodes. We shine mid-infrared light ($\hbar\omega\simeq120$ meV) under normal incidence onto the sample and collect the photocurrent without applying any electrical bias. \textbf{c,} The CPGE currents of bilayer and monolayer WTe$_2$ samples as a function of the displacement field under similar conditions (device geometry, flake size, temperature, etc). \textbf{d,} By flowing an AC current ($I_{\textrm{exc}}^{\omega}//\hat{a}$, frequency $\omega=177.77$ Hz) between electrodes $7-8$, we detect the nonlinear voltages $V^{2\omega}$ along both the longitudinal and transverse directions.  \textbf{e,} $V^{2\omega}_{baa}$ at different gate voltages. $V^{2\omega}_{baa}$ denotes the $V^{2\omega}$ along $\hat{b}$ in response to $I^{\omega}//\hat{a}$. The solid lines in panels (\textbf{d,e}) are quadratic fits to the data. Data in panels (\textbf{b-d}) are collected at $T=50$ K; Data in panel (\textbf{e}) are collected at $T = 100$ K.}
\label{Fig2}
\end{figure*}

\begin{figure*}[t]
\includegraphics[width=16cm]{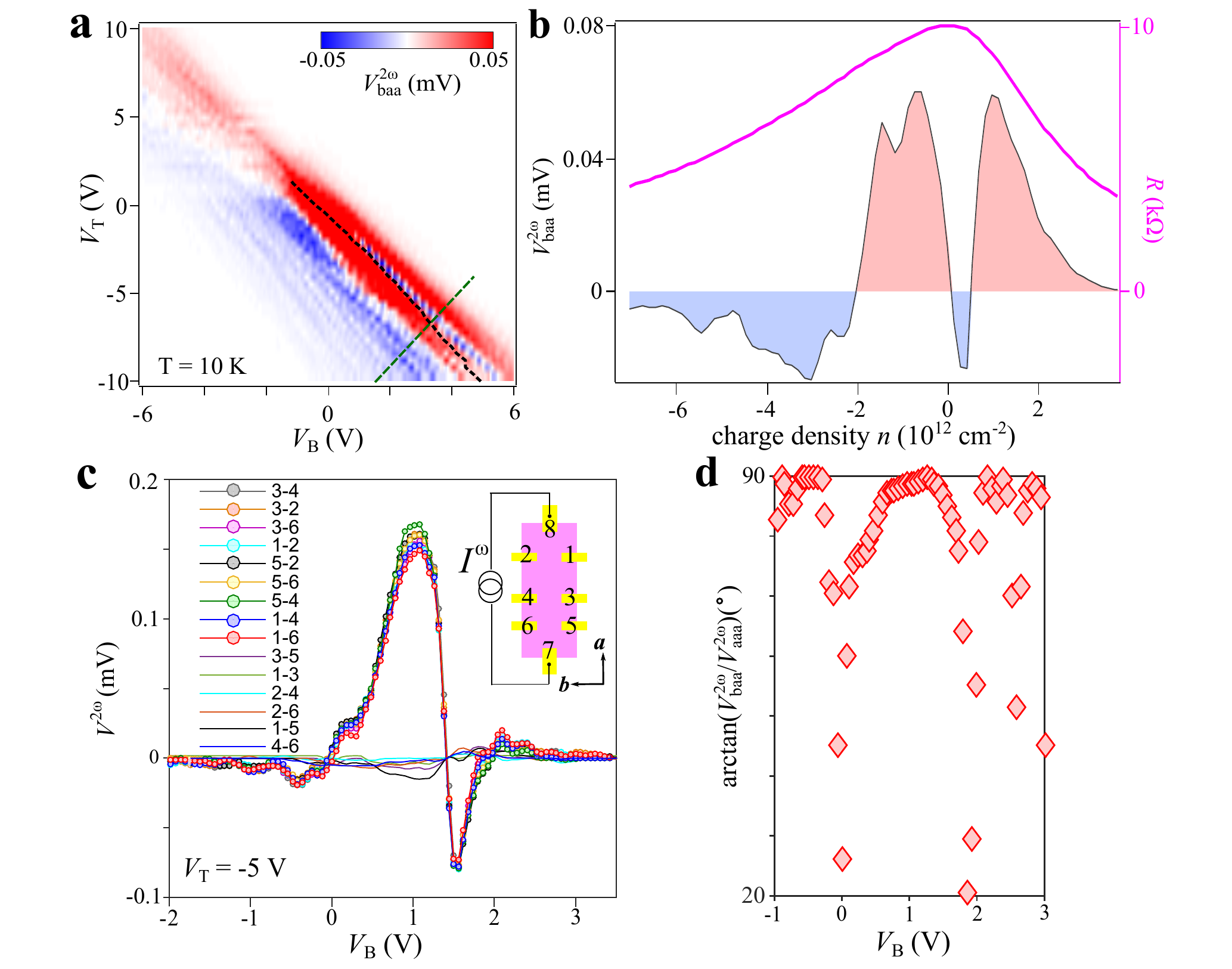}
\caption{{\bf Systematic studies of the nonlinear Hall effect in bilayer WTe$_2$.} \textbf{a,} Gate map of the $V^{2\omega}_{baa}$ at $T=10$ K. A current $I^{\omega}_{\textrm{exc}}=1$ $\mu$A is applied for all gate voltages. The black dashed line traces the ($V_{\textrm{T}}, V_{\textrm{B}}$) locations that correspond to the maximum of the resistance (Fig.~\ref{Fig1}\textbf{i}) at every $D$-field value (see SI.II.1). The green line denote a trajectory in the gate map where the charge density is varied at a constant $D$-field. \textbf{b,} $V^{2\omega}_{baa}$ and resistance as a function of charge density along the green dashed line in panel (\textbf{a}). \textbf{c,} $V^{2\omega}$ measured between many combinations of electrodes while keeping $I^{\omega}//\hat{a}$. The transverse (longitudinal) signals are shown by lines with (without) circles. \textbf{d,} The nonlinear Hall angle $\arctan(\frac{V^{2\omega}_{baa}}{V^{2\omega}_{aaa}})$ determined from the data in panel (\textbf{c}).}
\label{Fig3}
\end{figure*}

\begin{figure*}[t]
\includegraphics[width=17cm]{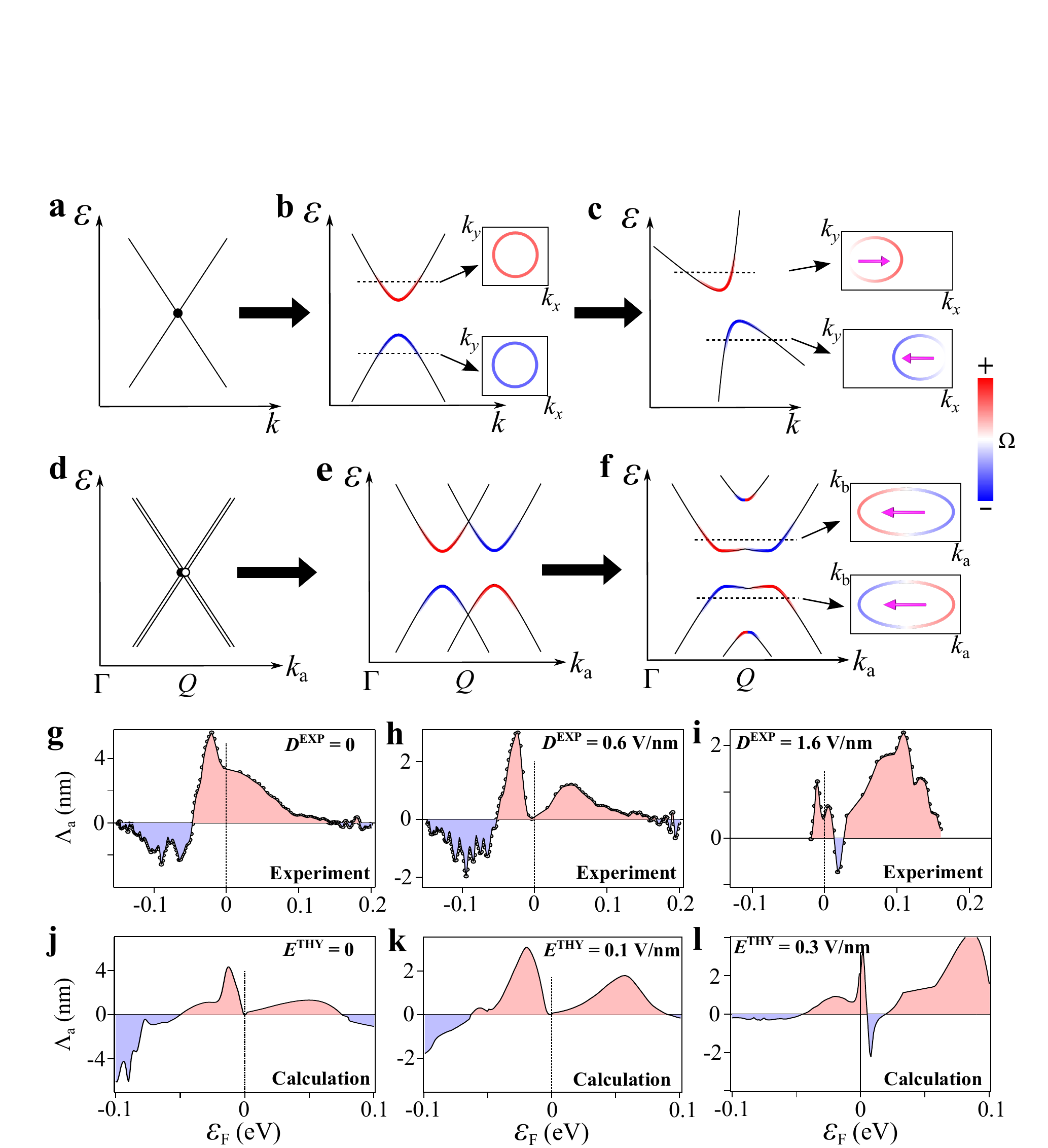}
\caption{{\bf Large Berry curvature dipole arising from layer-polarized Dirac fermions in bilayer WTe$_2$.} \textbf{a,} A massless Dirac fermion does not have any Berry curvature. \textbf{b,} Gapping it out by breaking inversion symmetry leads to large Berry curvature $\Omega$ near the gap edges. \textbf{c,} Tilting the Dirac cone allows the left- and right-movers to have different $\Omega$ and different $\frac{\partial\varepsilon}{\partial k_\alpha}$, leading to a nonzero $\vec{\Lambda}$, which has opposite signs for the conduction and valence bands (See Eq.~\ref{BCD}). \textbf{d,} In the absence of inter-layer coupling and SOC, the $Q$ point of bilayer WTe$_2$ features a pair of degenerate, massless Dirac fermions (one from each layer) of opposite chirality. }
\label{Fig4}
\end{figure*}
\addtocounter{figure}{-1}
\begin{figure*}[t!]
\caption{\textbf{e,} We now turn on inter-layer coupling. The two Dirac fermions are both gapped because of the explicit breaking of inversion symmetry due to the inter-layer coupling. For a gapped Dirac fermion, the conduction and valence bands have opposite Berry curvature. The absolute sign of the Berry curvature depends on the chirality of the Dirac fermion. The two Dirac fermions are shifted with respect to each other. \textbf{f,} The inter-layer coupling further opens up gaps wherever the Dirac fermions cross. As a result, we arrive at a quite dramatic scenario where the left- and right-movers have the opposite Berry curvatures, which therefore surely leads to a large $\vec{\Lambda}$ (see Eq.~\ref{BCD}). $\vec{\Lambda}$ has the same sign for the conduction and valence bands, which is a unique signature of the asymmetrically coupled Dirac fermions in bilayer WTe$_2$. \textbf{g-i,} Experimentally obtained Berry curvature dipole $\Lambda_a$ based on the $V^{2\omega}_{baa}$ data in Fig.~\ref{Fig3}\textbf{a} (see SI.VII for details) at three different displacement fields. 
Strictly speaking, the data allow us to determine $\Lambda_a$ as a function of the charge density $n$ at a given displacement field $D$ (see SI. VII). We further relate the charge density $n$ to the chemical potential $\varepsilon_{\textrm{F}}$ according to first-principles calculations (SI. VII). \textbf{j-l,} Theoretically calculated $\Lambda_a$ at three different displacement fields. The theoretically applied electric field ($E^{\textrm{THY}}$) here can be related to the displacement field by $D^{\textrm{THY}}=\epsilon^{\textrm{WTe}_2}E^{\textrm{THY}}$. }
\end{figure*}

\setcounter{figure}{0}
\renewcommand{\figurename}{\textbf{Extended Data Fig.}}

\clearpage
\begin{figure*}[t]
\includegraphics[width=17cm]{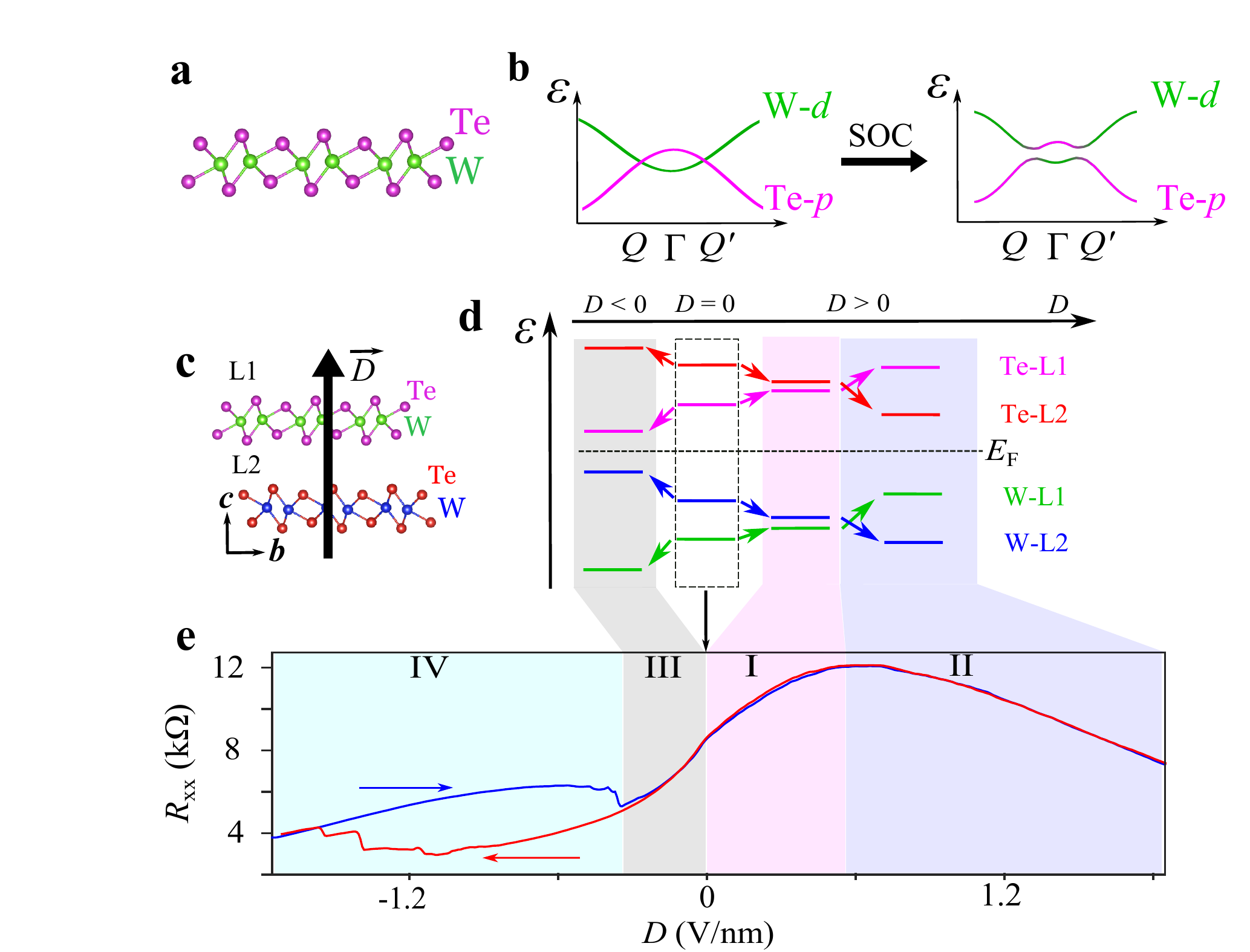}
\caption{{\bf Asymmetric inter-layer coupling.} \textbf{a,b,} In monolayer WTe$_2$, the low energy physics is characterized by a W-$5d$ band and a Te-$5p$ band. Due to the band inversion near $\Gamma$, a Dirac fermion is formed at the $Q$($Q'$) point. Without SOC, the Dirac fermion is gapless because of the inversion symmetry of the $1T'$ crystal structure \cite{muechler2016}. The inclusion of SOC opens up a gap, leading to a quantum spin Hall state. \textbf{c,d,} Bilayer WTe$_2$ is described by a pair of Dirac fermions at the $Q$ point, which consist of four bands. To isolate the essential physical picture for the resistance behavior, we ignore the wave-vector $k$ and only look at the $\Gamma$ point. The four bands further simplify into four energy levels, namely, W-$5d$-layer1 (W-L1), W-$5d$-layer2 (W-L2), Te-$5p$-layer1 (Te-L1) and Te-$5p$-layer2 (Te-L2). Due to the asymmetric inter-layer coupling, orbitals from opposite layers gain different on-site potentials, which lifts the layer degeneracy. We note that this coupling is only allowed when the system lacks any symmetry that relates $\hat{c}$ to $-\hat{c}$ (see SI.II.3). An external $D$-field along one direction enhances the layer splitting, whereas a $D$-field along the opposite direction reduces and eventually reverses the splitting. The data in panel (\textbf{e}) can be nicely explained by correlating the resistance with the global band gap in panel (\textbf{d}). The asymmetric coupling picture is confirmed by our first-principle calculation in Extended Data Fig.~\ref{EF2}. }
\label{EF1}
\end{figure*}

\clearpage
\begin{figure*}[t]
\includegraphics[width=17cm]{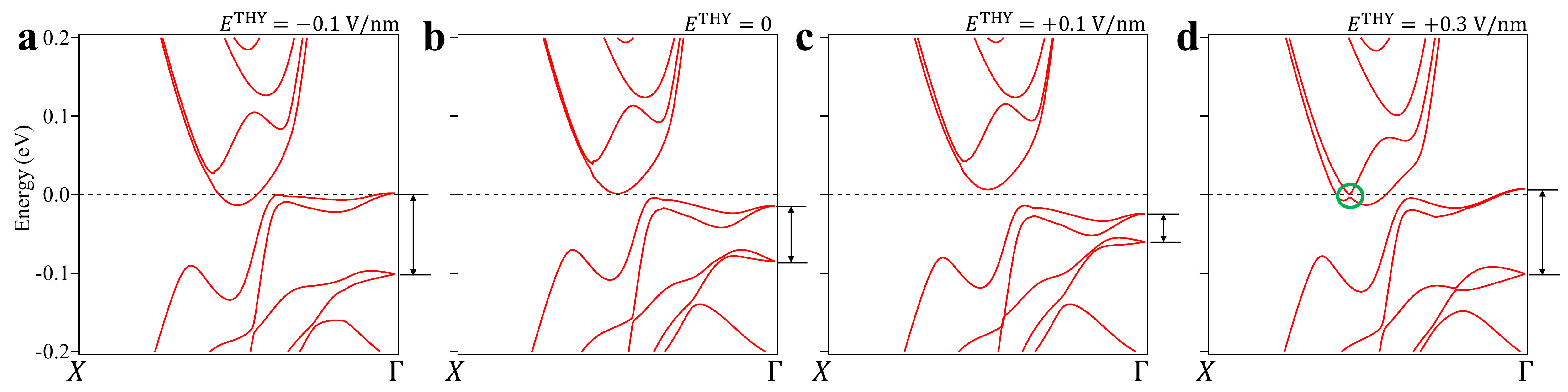}
\caption{{\bf Band structure of bilayer WTe$_2$ as a function of the external electric field.} The evolution of the valence band splitting at $\Gamma$ (denoted by the black arrows) is consistent with the picture in Extended Data Fig.~\ref{EF1}. With positive $E$-fields, the global band gap increases then decreases; With negative $E$-fields, the global band gap decreases. The evolution of the global band gap as a function of the $E$-field agrees with the picture in Extended Data Fig.~\ref{EF1}. The green circle highlights the anti-crossing (a massive Dirac fermion) formed between the lowest two conduction bands at $D>0$. Such a massive Dirac cone with a small gap ($\simeq3$) meV leads to large Berry curvatures near the gap edges, which are responsible for the additional sharp features in the $V_{baa}^{2\omega}$ data (see SI. V).  }
\label{EF2}
\end{figure*}

\vspace{1cm}
\begin{figure*}[h]
\includegraphics[width=15cm]{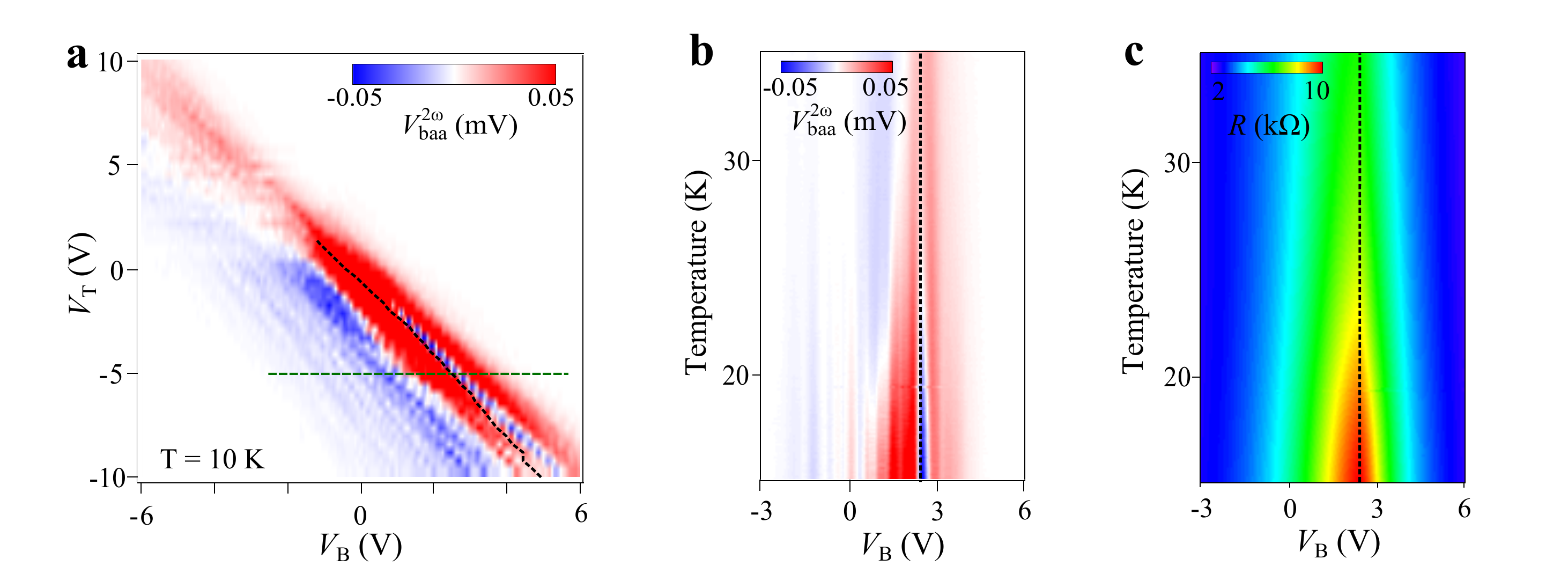}
\caption{{\bf Gate and temperature dependence of the nonlinear Hall voltage and resistance.} \textbf{a,} Gate map of the nonlinear Hall voltage $V^{2\omega}_{baa}$. \textbf{b,} Temperature dependence of the nonlinear Hall voltage $V^{2\omega}_{baa}$ as a function of the back gate voltage $V_{\textrm{B}}$ at $V_{\textrm{T}}=-5$ V (noted by the green dashed line) in panel (\textbf{a}). \textbf{c,} Same as panel (\textbf{b}) but for the resistance. }
\label{EF3}
\end{figure*}
\end{document}